\documentclass[twocolumn,prl,superscriptaddress, floatfix]{revtex4-1}

\usepackage{setspace, graphicx, color, amssymb, amsmath, subfigure, verbatim, ulem}
\usepackage[all]{xy}
\pdfoutput=1

\newtheorem{theorem}{Theorem}

\ifx\pdfoutput\undefined
	\usepackage[hypertex]{hyperref}
\else
	\usepackage[pdftex, hypertexnames=false]{hyperref}
\fi

    \def \R {{\mathbb{R}}}

   \def \Z {{\mathbb{Z}}}
    \def \T {{\mathbb{T}}}
    \def \e {{\varepsilon}}

    \def \o {{\theta}}
    
    \def \P {{\Phi}}

    \def \z {{y_2}}

    \def \F {{\mathcal{F}}}
    \def \O {{\mathcal{O}}}
    \def \D {{\Delta}}

    \def \l {{\lambda}}

    \def \s {{\sigma}}
    
    \def \b {{\beta}}

    \def \n {{\eta}}

    \def \z {{\zeta}}
    
    \def \Rs{{R_{\text{spike}}}}
    \def \proj {{p_{i,\zeta}^t}}

\newcommand{\beq}{\begin{equation}}
\newcommand{\eeq}{\end{equation}}
\newcommand{\beqr}{\begin{eqnarray}}
\newcommand{\eeqr}{\end{eqnarray}}
\newcommand{\beqrn}{\begin{eqnarray*}}
\newcommand{\eeqrn}{\end{eqnarray*}}
\newcommand{\beqn}{\begin{equation*}}
\newcommand{\eeqn}{\end{equation*}}
\newcommand{\bei}{\begin{itemize}}
\newcommand{\beii}{\begin{itemize} \item}
\newcommand{\eei}{\end{itemize}}
\newcommand{\bmei}{\begin{itemize} \compactlist}
\newcommand{\emei}{\end{itemize}}
\newcommand{\ben}{\begin{enumerate}}
\newcommand{\een}{\end{enumerate}}
\newcommand{\bes}{\begin{small}}
\newcommand{\ees}{\end{small}}
\newcommand{\bec}{\begin{center}}
\newcommand{\eec}{\end{center}}

\newcommand{\tht}{\theta}

\usepackage{soul}

\begin{document}

\begin{abstract} 

Biological information processing is often carried out by complex networks of interconnected dynamical units.  A basic question about such networks is that of {\it reliability:} if the same signal is presented many times with the network in different initial states, will the system entrain to the signal in a repeatable way?  Reliability is of particular interest in neuroscience, where large, complex networks of excitatory and inhibitory cells are ubiquitous.  These networks are known to autonomously produce strongly chaotic dynamics --- an obvious threat to reliability.  Here, we show that such chaos persists in the presence of weak and strong stimuli, but that even in the presence of chaos, intermittent periods of highly reliable spiking often coexist with unreliable activity.
We elucidate the local dynamical mechanisms involved in this intermittent reliability, and investigate the relationship between this phenomenon and certain time-dependent attractors arising from the dynamics.  A conclusion is that chaotic dynamics do not have to be an obstacle to precise spike responses, a fact with implications for signal coding in large networks.
\end{abstract}

\title{Chaos and reliability in balanced spiking networks with temporal drive}
\author{Guillaume Lajoie$^1$, Kevin K.~Lin$^2$, Eric Shea-Brown}
\affiliation{University of Washington, Dept of Applied Mathematics ; $^2$ University of Arizona, Dept of Mathematics}
\date{\today}
\maketitle

\section{Introduction}
Information processing by complex networks of interconnected dynamical units occurs in biological systems on a range of scales, from intracellular genetic circuits to nervous systems \cite{Eigen:1981p16930,Bialek:1991p6500}.  In any such system, a basic question is the {\it reliability} of the system i.e., the reproducibility of a system's output when presented with the same driving signal but with different initial system states.  This is because the degree to which a network is reliable constrains how --- and possibly how much --- information can be encoded in the network's dynamics.  This concept is of particular interest in computational neuroscience, where the degree of a network's reliability determines the precision (or lack thereof) with which it maps sensory and internal stimuli onto temporal spike patterns.
Analogous phenomena arise in a variety of physical and engineered systems, including coupled lasers~\cite{Uchida:2004p16812} (where it is known as ``consistency'') and ``generalized synchronization'' of coupled chaotic systems~\cite{Rulkov:1995p16814}.

The phenomenon of reliability is closely related to questions of dynamical stability, and in general whether a network is reliable reflects a combination of factors, including the dynamics of its components, its overall architecture, and the type of stimulus it receives \cite{LinSY07a}.  Understanding the conditions and dynamical mechanisms that govern reliability in different classes of biological network models thus stands as a 
challenge in the study of networks of dynamical systems.
An ubiquitous and important class of neural networks are those with a balance of excitatory and inhibitory connections~\cite{Shu:2003p16590}.  Such {\it balanced networks} produce dynamics that match the irregular firing observed experimentally on the ``microscale" of single cells, and on the macroscale can exhibit a range of behaviors, including rapid and linear mean-field dynamics that could be beneficial for neural computation~\cite{Tomko:1974p16592,Noda:1970p16591,Vreeswijk:1998p14451,Sha+98,Monteforte:2010p11768}.  However, such balanced networks are known to produce strongly chaotic activity when they fire autonomously or with constant inputs~\cite{Vreeswijk:1998p14451,London:2010p10818, Monteforte:2010p11768}.  On the surface, this {may appear}  incompatible with reliable spiking, as small differences in initial conditions between trials may lead to very different responses.  
However, that the answer might be more subtle is suggested by a variety of results on the impact of temporally fluctuating inputs on chaotic dynamics~\cite{PhysRevLett.69.3717,Banerjee:2008p6505, Rajan:2010p7924,LinSY07a,Lin:2009p16577,LitwinKumar:2011p16597,Bazhenov:2005p16805}.

At a more technical level, because of the link between reliability and dynamical stability, many previous theoretical studies of reliability of single neurons and neuronal networks have focused  on the {\it maximum Lyapunov exponent} of the system as an indicator of reliability.  This is convenient because (i) exponents are easy to estimate numerically and, for certain special types of models, can be estimated analytically \cite{Vreeswijk:1998p14451, Ritt:2003p145,01, London:2010p10818,Pakdaman:2001p5880,01,LinSY07a}; and (ii) using a single summary statistic permits one to see, at a glance, the reliability properties of a system across different parameter values.  However, being a single statistic, the maximum Lyapunov exponent cannot capture all relevant aspects of the dynamics.  Indeed, the maximum exponent measures the rate of separation of trajectories in the most unstable phase space direction; other aspects of the dynamics are missed by this metric.
  Recently, attention has turned to the full Lyapunov spectrum.  In particular, \cite{Monteforte:2010p11768} compute this spectrum for balanced autonomously spiking neural networks, and suggest limitations on information transmission that result.

In this paper, we present a detailed numerical study and steps toward a qualitative theory of reliability in fluctuation-driven networks with balanced excitation and inhibition.  
One of our main findings is that even in the presence of strongly chaotic activity -- as characterized by positive Lyapunov exponents -- single cell responses can exhibit intermittent periods of sharp temporal precision, punctuated by periods of more diffuse, unreliable spiking.  We elucidate the local (meaning cell-to-cell) interactions involved in this intermittent reliability, and investigate the relationship between this phenomenon and certain time-dependent attractors arising from the dynamics (some geometric properties of which can be deduced from the Lyapunov spectrum).

\section{Model description}
We study a temporally driven network of $N=1000$ spiking neurons.  Each neuron is described by a phase variable $\tht_i \in S^1=\R/\Z$ whose dynamics follow the ``$\tht$-{\it neuron}" model~\cite{Ermentrout:1996p10447}. 
This model's spike generation in so-called ``Type I'' neurons and are equivalent to the ``quadratic integrate-and-fire'' (QIF) model after a change of coordinates (see \cite{Ermentrout:1996p10447,LathamRNN00} and the Appendix).  These models can also be formally derived from biophysical neuron models near ``saddle-node-on-invariant-circle'' bifurcations; the underlying ``normal form" dynamics~\cite{Ermentrout:1996p10447,Ermentrout:2010p10442} are found in many brain areas.  The $\theta$-neuron model is known to produce reliable responses to stimuli in isolation~\cite{LinSY07a, Ritt:2003p145}, cf.~\cite{Mainen:1995p16589,Bry+76}.  Thus, any unreliability or chaos that we find is purely a consequence of network interactions.

Coupling from neuron $j$ to neuron $i$ is determined by the weight matrix $A=\{a_{ij}\}$. $A$ is chosen randomly  as follows: each cell is either excitatory (i.e., all its out-going weights are $\geq0$) or inhibitory (all its out-going weights are $\leq0$), with 20\% of the cells $j$ being inhibitory and 80\% excitatory; we do not allow self-connections, so $a_{ii}=0$. Each neuron has {\it mean} in-degree $K=20$ from each population (excitatory and inhibitory) and the synaptic weights are
$\O(1/\sqrt{K})$ in accordance with the classical balanced-state network architecture~\cite{Vreeswijk:1998p14451}. 
We note that our results appear to be qualitatively robust to changes in $N$ and $K$, 
 but a detailed study of scaling limits is beyond the scope of this paper.

 \begin{figure}[b!]
\begin{center}
\includegraphics{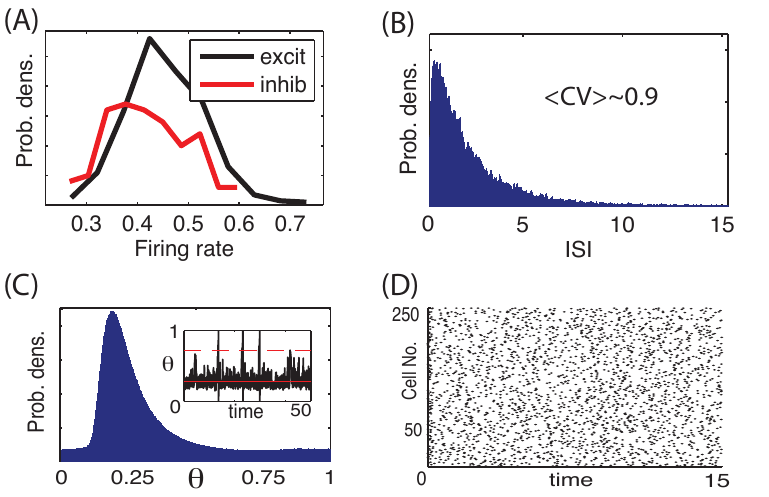}
\caption{(Color online) (A) Typical firing rate distributions for excitatory and inhibitory populations. (B) Typical inter-spike-interval (ISI) distribution of a single cell. The coefficient of variation (CV) is close to 1. (C) Invariant measure for an excitable cell ($\n<0$); inset: typical trajectory trace of an excitable cell where solid and dotted lines mark the stable and unstable fixed points. (D) Network raster plots for 250 randomly chosen cells.  For all panels, $\n=-0.5$, $\e=0.5$.}
\label{fig:general}
\end{center}
\end{figure}

A neuron $j$ is said to fire a spike when $\o_j(t)$ crosses $\o_j=1$; when this occurs, $\o_i$ is impacted via the coupling term $a_{ij}g(\o_j)$ where $g(\o)$ is a smooth ``bump" function with small support ($[-1/20,1/20]$) around $\o=0$ satisfying $\int_{0}^1g(\o)d\o=1$, meant to model the rapid rise and fall of a synaptic current (see Appendix for details). 
In addition to coupling interactions, each cell receives a stimulus $I_i(t)=\n+\e \z_i(t)$ where $\n$ represents a constant current and $\zeta_i(t)$  are aperiodic signals, modeled here (as in~\cite{Bry+76,Mainen:1995p16589, LinSY07a}) by ``frozen'' realizations of independent white noise processes, scaled by an amplitude parameter $\e$.   Note that the terms $\z_i(t)$ model external signals, not ``noise'' (i.e., driving terms that can vary between trials), though such terms can be easily added (as in \cite{Lin:2009p16577}).

 The $i^{\text{th}}$ neuron in the network is therefore described by the following stochastic differential equation (SDE):
\beq
\begin{split}
 d\o_i&=[ F(\o_i)+Z(\o_i)\left ( \n+\sum_j a_{ij}g(\o_j) \right )+\\
 & \frac{\e^2}{2} Z(\o_i)Z'(\o_i)] dt+\e Z(\o_i)\cdot dW_{i,t}
\end{split}
\label{stoch_tht}
\eeq
 where the intrinsic  dynamics $F(\o_i)=1+\cos(2\pi \o_i)$ and the stimulus response curve $Z(\o_i)=1-\cos(2 \pi \o_i)$ come directly from coordinate changes based on the original QIF equations (see Appendix and \cite{Ermentrout:1996p10447}). Here, $W_{i,t}$ is the independent Wiener process generating $\z_i(t)$; the $\e^2$ term is the It\^{o} correction from the coordinate change~\cite{Lindner:2003p16585}.  
Finally, 
$\n$ sets the intrinsic excitability of individual cells.  
For $\n<0$, there is a stable and an unstable fixed point, together representing resting and threshold potentials.  Thus (contrasting~\cite{Monteforte:2010p11768} where cells are intrinsically oscillatory), neurons are in the ``excitable regime," displaying  {\it fluctuation-driven} firing, as for many cortical neurons~\cite{Destexhe:2003p16747}.

In what follows, we focus on networks in this regime by fixing $\n=-0.5$, where cells spike due to temporal fluctuations in their inputs (both from external drive and network interactions) rather than being perturbed and coupled oscillators. We study the effect of the amplitude $\e$ of the external drive on the evoked dynamics.  Note that in the absence of such inputs, these networks do not produce sustained activity. 

Fig.~\ref{fig:general} illustrates that the general properties of the network dynamics, including a wide distribution of firing rates from cell to cell and highly irregular firing in individual cells, are consistent with many models of balanced-state networks in the literature, as well as general empirical observations from cortex~\cite{Noda:1970p16591, Tomko:1974p16592}. An additional such property is that our network's mean firing rate scales monotonically with $\n$ and $\e$ (data not shown), as in~\cite{Vreeswijk:1998p14451, Monteforte:2010p11768}. 

\section{Mathematical background}

For reliability questions, we are interested in the response of a network to a fixed input signal starting from different initial states.  Equivalently, we can imagine an {\it ensemble} of initial conditions all being driven simultaneously by the same signal $\z(t)~.$  If the system is reliable, then there should be a distinguished trajectory $\o(t)$ to which the ensemble converges.  In contrast, an unreliable network will lack such an attracting solution, as dynamical mechanisms conspire to keep trajectories separated. To put these ideas on a precise mathematical footing, it is useful to treat our SDE~(\ref{stoch_tht}) as a {\it random dynamical system (RDS).} That is, we view the system as a nonautonomous ODE driven by a frozen realization of the Brownian process, and consider the action of the generated family of flow maps on phase space. In this section, we present a brief overview of RDS concepts and their meaning in the context of network reliability.

\subsection{Random dynamical systems framework}
 The model network described by~(\ref{stoch_tht}) is a SDE of the form 
\begin{equation}
  dx_t=a(x_t)dt+\sum_{i=1}^N b(x_t)\cdot dW^i_t
  \label{rds_gen}
\end{equation}
whose domain is the $N$-dimensional torus $\T^N$ and $W_t^i$ are standard Brownian motions.  We assume throughout that the Fokker-Planck equation associated with~(\ref{rds_gen}) has a unique, smooth steady state solution $\mu$. Since we are interested in the time evolution of an ensemble of initial conditions driven by a {\it single, fixed realization} $\z$ generated by $\{W_t^i\}_i~,$ this can be done by considering the {\it stochastic flow maps} defined by the SDE, i.e., the solution maps of the SDE.  More precisely, this is a family of maps $\Psi_{t_1,t_2;\z}$ such that $\Psi_{t_1,t_2;\z}(x_{t_1})=x_{t_2}$ where $x_t$ is the solution of~(\ref{rds_gen}) given $\z$. If $a(x)$ and $b(x)$ from ~(\ref{rds_gen}) are sufficiently smooth, it has been shown (see, e.g.~\cite{kunita}) that the maps $\Psi_{t_1,t_2;\z}$ are well defined, smooth with smooth inverse (i.e., are diffeomorphisms), and are independent over disjoint time intervals $[t_1,t_2]$.

RDS theory studies the action of these random maps on the state space.  The object from RDS theory most relevant to questions of reliability is the {\it sample distribution} $\mu^t_\z$~, defined here as
\beq
\mu_\z^t=\lim_{s \to -\infty}(\Psi_{s,t;\z})_*\mu_{init}~,
\label{transfer}
\eeq
where $(\Psi_{s,t;\z})_*$  denotes the propagator associated with the flow $\Psi_{s,t;\z}$, i.e., it is the linear operator transporting probability distributions from time $s$ to time $t$ by the flow $\Psi_{s,t;\z}$, and $\mu_{init}$ is the initial probability distribution of the ensemble.  

The definition above has the following interpretation: suppose the system was prepared in the distant past so that it has a random initial condition (where ``random'' means ``having distribution $\mu_{init}$").  Then $\mu_\z^t$ is precisely the distribution of all possible states at time $t$, after the ensemble has been subjected to a given stimulus $\z(t)$ for a sufficiently long time (how long is ``sufficient'' is system-dependent; the limit in the definition sidesteps that question).
So if $\mu_\z^t$ were localized in phase space (i.e., if its support has relatively small diameter), then its state at at time $t$ is essentially determined solely by the stimulus up to that point, i.e., its response at time $t$ is reliable.  In contrast, if  $\mu_\z^t$ were not localized, then the response is unreliable in the sense that the system's initial condition has a measurable effect on its state at time $t$. Note that $\mu_\z^t$ depends on both $\z$ and the time $t$: as time goes by, the system receives more inputs, and $\mu_\z^t$ continues to evolve; it is easy to see that $(\Psi_{t_1,t_2;\z})_*\mu_\z^{t_1} = \mu_\z^{t_2}~.$ In general, we expect $\mu_\z^t$ to be essentially independent of the specific choice of  $\mu_{init}$, so long as $\mu_{init}$ is given by a sufficiently smooth probability density, e.g., the uniform distribution on $\T^N~.$

\subsection{Linear stability implies reliability}

Not surprisigly, the reliability of a system is related to its {\it dynamical stability.}  This link can be made precise via the {\it Lyapunov exponents} $\l_1 \geq \l_2 \geq ... \geq \l_N$ of the stochastic flow.  As in the deterministic case, these exponents measure the rate of separation of nearby trajectories; for a ``typical'' trajectory, we expect a small perturbation $\delta{x}_t$ to grow or contract like $|\delta{x}_t|\sim e^{\lambda_1 t}$ over sufficiently long timescales.  Note that under very general conditions, the exponents are deterministic, i.e., they depend only on system parameters but not on the specific realization of the input $\z$~\cite{kifer}.  Moreover, consistent with the findings in \cite{Monteforte:2010p11768,Lin:2009p16577}, we have observed that the exponents for our models are insensitive to specific realizations of the coupling matrix $A$ (see Appendix), so that they are truly functions of the system parameters.

One link between exponents and $\mu_\z^t$ is the following theorem:
\begin{theorem}[Le Jan; Baxendale~\cite{LeJan:1987p129,Braxendale:1991p11730}]
 If $\lambda_1<0$ and a number of nondegeneracy conditions are satisfied \cite{Braxendale:1991p11730}, then $\mu_\z^t$ is a {\it random sink}, i.e., $\mu_\z^t(x) = \delta(x-x_t)$ where $x_t$ is a solution of the SDE.
\end{theorem}
Theorem 1 states that under broad conditions, an ensemble of
trajectories described by a smooth initial density will collapse
toward a single, distinguished trajectory.
For this reason,  $\lambda_1<0$ is often associated with reliability.

A second, complementary theorem covers the case $\lambda_1>0$.

\begin{theorem}[Ledrappier and Young~\cite{Led+88}]
  If $\lambda_1>0$, then $\mu_\z^t$ is a {\it random Sinai-Ruelle-Bowen (SRB) measure}.
\end{theorem}
\noindent SRB measures are concepts that originally arose in the theory of deterministic, dissipative chaotic systems~\cite{RevModPhys.57.617,Young:2002p10781}.  They are singular invariant probability distributions supported on a ``strange attractor.''  Such attractors necessarily have zero phase volume because of dissipation; nevertheless, SRB measures capture the statistical properties of a set of trajectories of positive phase volume (i.e., the strange attractor has a nontrivial basin of attraction).  They are the ``smoothest'' invariant probability distributions for such systems in that they have smooth conditional densities along unstable (expanding) phase directions.  Indeed, locally they typically consist of the cartesian products of smooth manifolds with Cantor-like fractal sets; the tangent spaces $E_{u,\z}(x)$ to these smooth ``leaves'' are invariant in the sense that $D\Psi_{s,t;\z}(x_s)\cdot E_{u,\z}(x_s) = E_{u,\z}(x_t)$, where $D\Psi_{s,t;\z}(x)$ denotes the Jacobian of the flow map at $x$.  Moreover, these subspaces are
readily computable as a by-product of estimating Lyapunov exponents (see Appendix).

Random SRB measures share many of the same properties as SRB measures in the deterministic setting, but are time-dependent.  While in principle they may be confined to small regions of phase space at all times, this is typically not the case for the systems we study here.  A positive $\lambda_1$ is thus often associated with unreliability, and the terms ``chaotic'' and ``unreliable'' are often used interchangeably. (Random SRB measures have also been used to model the distribution of ``pond scum''; in that context they are known as ``snapshot attractors'' \cite{Namenson:1996p16601}.)

Although the SRB measure $\mu_\z^t$ evolves with time, it possesses some time-invariant properties because (after transients) it describes processes that are statistically stationary in time.
Among these is the dimension of the underlying attractor; another is the number of unstable directions, i.e., the number of positive Lyapunov exponents, which give the dimension of the unstable manifolds of the attractor.
The latter will be useful in what follows; we denote it by $M_\l~.$

\smallskip

To summarize, these two theorems allow us to reach global conclusions on the structure of random attractors (singular or extended) using only the maximum Lyapunov exponent $\lambda_1$, a measure of linear stability.  This has a number of consequences in what follows: first, because $\lambda_1$ is a single summary statistic determined only by system parameters (and not specific input or network realizations), it allows us to see quickly the reliability properties of a system across different parameters.  Second, unlike other measures of reliability, $\lambda_1$ can be computed easily in numerical studies by simulating single trials (as opposed to multiple repeated trials).
However, $\l_1$ can only tell us about reliability properties in an asymptotic sense (i.e., on sufficiently long timescales), and only about the dynamics in the  fastest expanding directions.  As we shall see later,  the reliability properties of our networks reflect the geometric properties of their SRB measures beyond those captured by $\lambda_1$ alone.

\section{Maximum Lyapunov exponents and asymptotic reliability}

In line with previous studies~\cite{Vreeswijk:1998p14451,London:2010p10818, Monteforte:2010p11768,Lin:2009p16577,Ritt:2003p145,Pakdaman:2001p5880,01,LinSY07a}, 
we say that a network is {\it asymptotically reliable} if $\l_1<0$ and {\it asymptotically unreliable} if $\l_1>0$.  In principle, even when $\l_1<0$, distinct trajectories could take very long times to converge to the random sink. However, we note that for all asymptotically reliable networks we considered, convergence is typically achieved within about 10 time units. For the remainder of the paper, we will concentrate on ``steady state" dynamics and we adopt the point of view that ensembles of solutions for all systems considered were initiated in the sufficiently distant past.  The question of transient times, although very interesting, falls outside of the scope of this paper. 
\begin{figure}
\begin{center}
\includegraphics{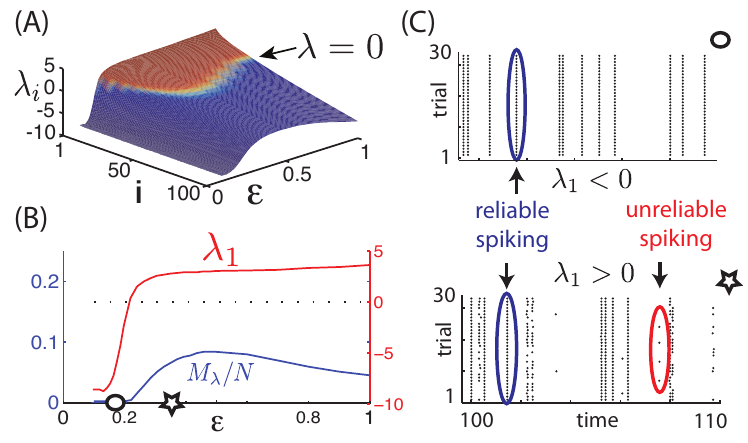}
\caption{(Color online) (A) First 100 Lyapunov exponents of network with fixed parameters as in Fig.~\ref{fig:general}, as a function of $\e$. (B) Plot of $\l_1$ (right scale), $M_\l/N$: the fraction of $\l_i>0$ (left scale) vs $\e$. (C) Raster plots show example spike times of an arbitrarily chosen cell in the network on 30 distinct trials, initialized with random ICs. Circle and star markers indicate $\e$ values of 0.18 and 0.5, respectively, shown in panel (B).  For all panels, $\n=-0.5$.}
\label{fig:color}
\end{center}
\end{figure}

We begin by studying the dependence of the $\l_i's$ 
on the input amplitude $\e$. Even in simple and low-dimensional, autonomous systems, analytical calculations of $\l_i$'s often prove to be very difficult if not impossible.  We therefore numerically compute (see Appendix for details) the Lyapunov spectra of our network for various values of input drive amplitude $\e$. Figure~\ref{fig:color} (A) shows the first 100 Lyapunov exponents of these spectra.  This demonstrates that, at intermediate values of $\e$, there are several positive Lyapunov exponents ($M_\l$), and that the trend in this number is nonmonotonic in $\e$.  Panel (B) gives another view of this phenomenon, as well as the dependence of $\l_1$ on $\e$.  In particular, for sufficiently small $\e$, the networks produce a negative $\l_1$. 

We note that for very small fluctuations ($\e<0.1$), the network rarely spikes and $\l_1$ is close to the real part of the largest eigenvalue associated with the stable fixed point of a single cell's vector field. As $\e$ increases, there is a small region ($0.1 < \e < 0.2$) where sustained network activity coexists with $\l_1<0$. However, as $\e$ increases further, there is a rapid transition to a positive $\l_1$, indicating chaotic network dynamics and thus asymptotic unreliability. Consistent with RDS theory, the transition to $\l_1>0$ is accompanied by the emergence of a random attractor with nontrivial unstable manifolds.

Since the networks we study are randomly connected and each cell is nearly identical, the underlying dynamics are fairly stereotypical from cell to cell.  This enables us to focus on a randomly chosen cell for illustrative purposes and further analysis. Figure~\ref{fig:color} (C) shows two sample raster plots where the spike times of a single cell from 30 distinct trials (initiated at randomly sampled ICs) are plotted. The top plot is produced from an asymptotically reliable system ($\l_1<0$) and as expected, every spike is perfectly reproduced on all trials. In the bottom plot, where $\l_1>0$, the spike times are clearly unreliable across different trials, as RDS theory predicts. For the remainder of this paper, we routinely refer to the parameter sets used in Fig~\ref{fig:color} (C) as testbeds for stable and chaotic networks respectively, and make use of them for illustrative purposes (see caption of Fig~\ref{fig:color} for details).

Finally, spike trains from the chaotic network also show an interesting phenomenon: there are many moments where spike times align across trials, i.e., the system is (temporarily) reliable.  We now investigate this phenomenon.

\section{Single-cell reliability}

Let us define the $i^{\text{{\it th}}}$ {\it neural direction} as the state space of the $i$th cell, which we identify with a circle $S^1~.$  
The degree of reliability of the $i$th cell is given by the corresponding marginal distribution, i.e., we define a projection $\pi_i(\theta_1,\cdots,\theta_N) = \theta_i$, and denote the corresponding projected single-cell distribution by $\proj(\o_i) \equiv \pi_i\mu_\z^t(\o)~.$  Note that when $\l_1>0$, we expect $\proj$ to be nonsingular, i.e., corresponds to a smooth probability density function (though it may be more or less concentrated); an exception is when the random attractor is aligned in such a way that it projects to a point onto the $i$th direction. If $\proj$ is singular at time $t$, then the state of cell $i$ is reproducible across trials at time $t$; geometrically, trajectories from distinct trials are perfectly aligned along the $i^{\text{th}}$ neural direction.  On the other hand, if $\proj$ has a broad density on $S^1$, then the state of cell $i$ at time $t$ can vary greatly across trials, and the $i$th components of distinct trajectories are separated.

\begin{figure*}
\begin{center}
\includegraphics{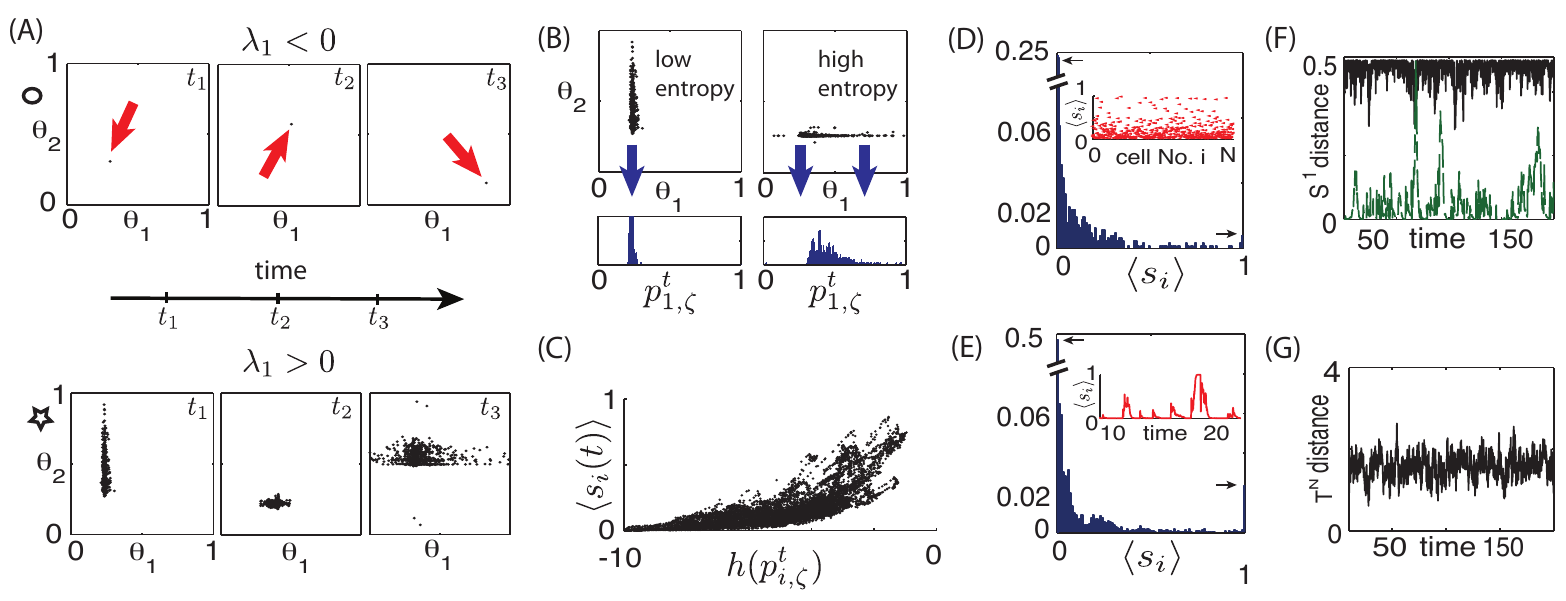}
\caption{(Color online) (A) Snapshots of 1000 trajectories projected in two randomly chosen neural directions ($\o_1$,$\o_2$) at three distinct times. Upper and lower rows with the same parameters as in Fig~\ref{fig:color} (C) and show a random sink and random strange attractor respectively.  (B) Projections of the sample measure $\mu_\z^t$ onto the $\o_1$ neural direction at distinct moments. (C) Scatter plot of average support score $\langle s_i(t) \rangle$ vs. entropy of projected measure $h(\proj)$ sampled over 2000 time points and 30 distinct cells. (D) Example histogram of $\langle s_i \rangle$ sampled across all cells in the network at a randomly chosen moment in time. Inset: snapshot of $\langle s_i \rangle$ vs. cell number $i$. (E) Example histogram of $\langle s_i(t) \rangle$ sampled across 2000 time points from a randomly chosen cell. Inset: sample time trace of $\langle s_i(t) \rangle$ vs. time. (F) and (G) Time evolution of distance between two distinct trajectories $\o^1(t)$, $\o^2(t)$ (F) Green dashed (bottom): $\|\o^1_i(t)-\o^2_i(t)\|_{S^1}$ in a randomly chosen $\o_i$ direction. Black solid (top): $\max_j\{ \|\o^1_j(t)-\o^2_j(t)\|_{S^1}\}$. (G) $\|\o^1(t)-\o^2(t)\|_{\T^N}$. For all panels except A (top), network parameters: $\n=-0.5$, $\e=0.5$ with $\l_1 \approx 2.5$.}
\label{fig:proj}
\end{center}
\end{figure*}

This is illustrated in Fig~\ref{fig:proj}(A) where snapshots of 1000 randomly initialized trajectories are projected onto ($\o_1$, $\o_2$)-coordinates at distinct times $t_1<t_2<t_3$. The upper snapshots are taken from an asymptotically reliable system ($\l_1<0$) where $\mu_\z^t$ is singular and supported on a single point (random sink) which evolves on $\T^N$ according to $\z(t)$. The bottom snapshots are taken from the $\l_1>0$ regime and clearly show that distinct trajectories accumulate on ``clouds" that change shape with time.  These changes affect the spread of $\proj$.

Our next task is to relate the geometry of the random attractor to the qualitative properties of the  single-cell distributions $\proj$~. A convenient tool for  quantifying the latter is the differential entropy  $h(\proj)=-\int_{S^1}d \proj \log_2\proj $.
Recall that the differential entropy of a uniform distribution on $S^1$ is 0, and that the more negative $h$ is, the more singular a distribution. In our context, the more orthogonal the attractor is to the $i^{\text{th}}$ direction in $\T^N$, the lower is its projection entropy, as illustrated in Fig~\ref{fig:proj}(B). We emphasize again that the shape of $\proj$ is time-dependent and so is its entropy.

\subsection{Uncertainty in single cell responses}

We would like to predict $h(\proj)$ from properties of the underlying dynamics.  Our first step in doing so is to validate our intuition about the orientation  of $\mu_\z^t$.  Following and somewhat generalizing an approach of~\cite{Monteforte:2010p11768}, we use a quantity which we call the {\it support score} $s_i(t)$ to represent the contribution of a neural direction to the unstable directions of the strange attractor at time $t$. 

We first define this quantity locally for a single trajectory $\o(t)$.  For this trajectory, we expect that there exists a decomposition of the tangent space into stable (contracting) and unstable (expanding) invariant subspaces: $E_{s,\z}(\o(t))$ and $E_{u,\z}(\o(t))$. Since the dimension of $E_{u,\z}(\o(t))$ must be $M_\l$, let $\{v_1, v_2,...,v_{M_\l} \}$ be an orthonormal basis for the unstable subspace at time $t$ (i.e. $v_i \in \R^N$). 
We define cell $i$'s {\it support score} as
\beq
s_i(t)=\|Vr_i\|
\label{support}
\eeq
where $V$ is the $M_\l \times N$ matrix with $v_i$'s as rows and $r_i$  is the ($N$-dimensional) unit vector in the $i^{\text{th}}$ direction. Note that $0 \leq s_i(t) \leq 1$, and that $s_i$ measures the absolute value of the cosine of the angle between the neural and unstable direction. Thus, $s_i$ represents the extent to which the  $i^{\text{th}}$ direction contributes to state space expansion. The vectors $\{v_1, v_2,...,v_{M_\l}\}$ are computed simultaneously with the $\l_i$'s (see numerical methods in Appendix). 

In order to use the support score to quantify the orientation of the attractor, we need to extend
the definition above, which is for a single trajectory, to an ensemble of trajectories governed by $\mu_\z^t$. 
However, $s_i(t)$ could greatly vary depending on which trajectory we choose --- as we might expect if $\mu_\z^t$ consisted of complex folded structures.
 Our numerical simulations show that this variation is limited in our networks: the typical variance of an ensemble of $s_i(t)$ values across an ensemble of trajectories with randomly chosen initial conditions is $\O(10^{-2})$ (for a fixed cell $i$ and a fixed time $t$). This suggests that unstable tangent spaces about many trajectories are similarly aligned.
Therefore, we extend the idea of support score to $\proj$ by taking the average $\langle s_i(t) \rangle$ across $\mu_\z^t$. We numerically approximate this quantity by averaging over 1000 trajectories. As stated earlier, the behavior of all cells are statistically similar because the network is randomly coupled.  As a consequence, the quantities $\langle s_i(t) \rangle$ and $\proj$ do not depend sensitively on which $i$ is chosen.

Figure~\ref{fig:proj} (C) shows a scatter plot of $\langle s_i(t) \rangle$ vs. $h(\proj)$ for a representative network that is asymptotically unreliable.
This clearly shows that the contribution of a neural direction $i$ to state space expansion results in a higher entropy of the projected measure $\proj$. 
This phenomenon is robust across all values of $\e$ tested. Once again, we note that this correspondence is not automatic for any dynamical system:  there is no guaranteed relationship between the orientation of the unstable subspace and the entropy of the projected density.
For example, the restriction of $\mu_\z^t$ to unstable manifolds could be very localized, thus having low entropy for even perfectly aligned subspaces.  

\subsection{Temporal statistics}

Next, we inquire about the distributions of $\langle s_i(t) \rangle$ across time and neural directions. That is, again following~\cite{Monteforte:2010p11768}, we study the number of cells that significantly contribute to unstable directions at any moment as well as the time evolution of this participation for a given cell. 

Figure~\ref{fig:proj} (D) shows a typical distribution of support scores across all cells in the network at a fixed moment in time. The inset shows a trace of $\langle s_i \rangle$ across cells at that moment. The important fact is that this is distribution is very uneven across neurons, being strongly skewed towards low values of $\langle s_i \rangle$.  In panel (E) of the same figure, we see a typical distribution of support scores across time for a fixed cell. The inset shows a sample of the $\langle s_i(t) \rangle$ time trace for that cell. We emphasize that the uneven shape of these distributions implies that at any given moment in time, only a few cells significantly support expanding directions of the attractor and moreover, that the identity of these cells change as time evolves. A similar mechanism was reported for networks of autonomously oscillating cells~\cite{Monteforte:2010p11768}, although only the maximally expanding direction was used to compute $s_i(t)$. In both cases, neurons in the network essentially take turns participating in the state space expansion that is present in the chaotic dynamics.

This leads to trajectories that are unstable on long timescales ($\lambda_1>0$), yet alternate between periods of stability and instability in single neural directions on short timescales. To directly verify this, Fig.~\ref{fig:proj} (F) shows a sample time trace of $\|\o^1_i(t)-\o^2_i(t)\|_{S^1}$: the projection distance between two randomly initialized trajectories $\o^1(t)$ and $\o^2(t)$ in a single neural direction $i$. Also shown is $\max_j\{ \|\o^1_j(t)-\o^2_j(t)\|_{S^1}\}$: the maximal projection distance out of all neural directions. While the maximal $S^1$ distance is almost always close to its maximum 0.5, the two trajectories regularly collapse arbitrarily close along any given $S^1$-direction. This leads to a global separation $\|\o^1(t)-\o^2(t)\|_{\T^N}$ that is relatively stable in time (Fig.~\ref{fig:proj} (G)) yet produces temporary local convergence ($\|\cdot\|_{\T^N}$ refers to the geodesic distance on the flat $N$-torus, i.e., a cube $[0,1]^N$ with opposite faces identified). In what follows, we will see that this mechanism translates into spike trains that retain considerable temporal structure from trial to trial.

\section{Reliability of spike times} 

Thus far, we have been concerned in general with the separation of trajectories arising from distinct trials (i.e. different ICs but fixed input $\z(t)$). However, of relevance to the dynamical evolution of the network state are spike times: the only moments where distinct neural directions are effectively coupled. Indeed, coupling between cells of this network is restricted to a very small portions of state space, namely to a small interval around $\o_i=0 \sim1$ when a cell spikes (see Model section). This property is ubiquitous in  neural circuits and other pulse-coupled systems~\cite{HopHer95} 
and is central to the time-evolution of $\mu_\z^t$.

\subsection{Spike reliability captured by probability fluxes}

From the perspective of spiking, what matters is the time evolution of projected measures on $S^1$ in relation to the spiking boundary. This is captured by the {\it probability flux} of $\proj$ at $\o_i=0 \sim 1$: $\P_i(t)$. For our system, we can easily write down the equation for the flux since inputs to a given cell have no effect at the spiking phase (ie. $Z(0)=0$ in (\ref{stoch_tht})). From (\ref{stoch_tht}), $\frac{d\o_i}{dt}|_{\o_i=0}=2$ and we have $\P_i(t)=2\proj(0)$. 
We emphasize that this probability flux is associated with $\mu_\z^t$, and differs from the usual flux arising from the Fokker-Planck equation.
Here, the source of variability between trajectories leading to wider $\proj$ is due to chaotic network interactions, rather than from noise that differs from trial to trial.  Overall, $\P_i(t)$ is modulated by a complex interaction of the stimulus drive $\z(t)$, the vector field of the system itself, and ``diffusion" originating from chaos; as we have seen, the latter depends in a nontrivial way on the geometric structure of the underlying strange attractor.

\begin{figure*}
\begin{center}
\includegraphics{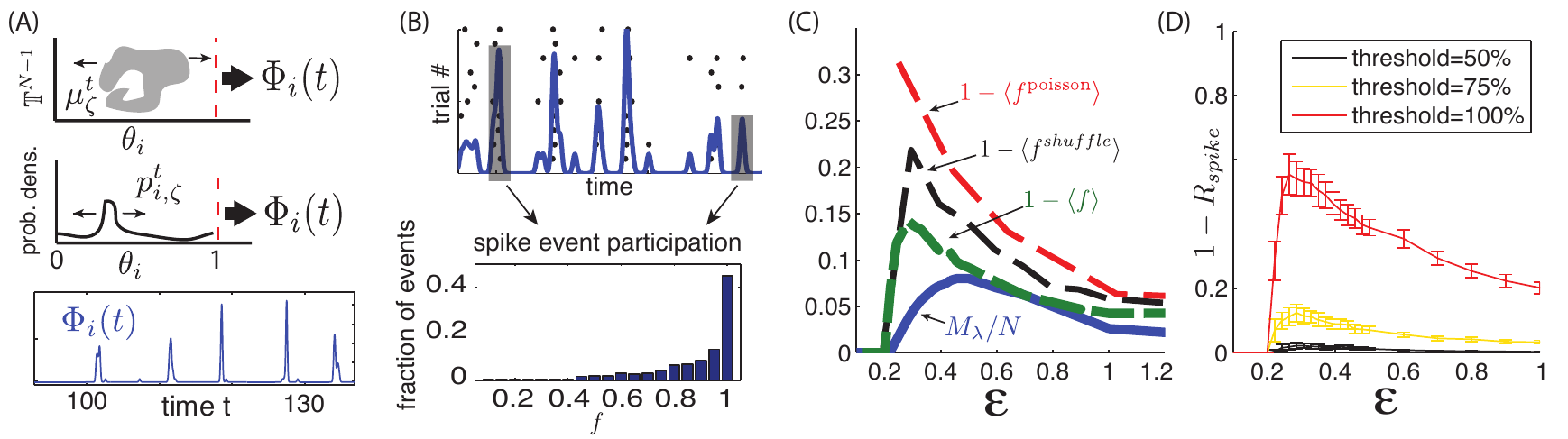}
\caption{(Colors online) (A) Top and middle: cartoon representations of the flux $\P_i(t)$. Bottom: sample $\P_i$ time trace for a randomly chosen cell approximated from 1000 trajectories. (B) Top: Illustration of spike event definition. Bottom: Distribution of spike event participation fraction $f$. (For (A) and (B): $\n=-0.5$, $\e=0.5$, $\l_1 \simeq 2.5$) (C) Curves of $1-\langle f \rangle$ (network), $1-\langle f^{shuffle} \rangle$ (single cell with shuffled input spike trains from networks simulations) and $1-\langle f^{poisson} \rangle$ (single cell with random poisson spike inputs) vs. $\e$. Also shown is the fraction of $\l_i>0$, $M_\l/N$ vs. $\e$. (D) Mean $1-\Rs$ vs. $\e$ curves for three threshold values. Error bars show one standard deviation of mean $\Rs$ across all cells in the network. (For (C) and (D): $\n=-0.5$)}
\label{fig:flux}
\label{fig:event}
\end{center}
\end{figure*}

In the limit of infinitely many trials, $\P_i(t)$ is exactly the normalized cross-trial spike time histogram, often referred to as the {\it peri-stimulus time histogram} (PSTH) in the neuroscience literature. A PSTH is obtained experimentally by repeatedly presenting the same stimulus to a neuron or neural system and recording the evoked spike times on each trial. Figure~\ref{fig:flux} (A) illustrates the time evolution of $\P_i$. Perfectly reliable spike times (repeated across all trials) are represented by a time $t^*$ such that for an open interval $U \ni t^*$,  $\P_i(t)|_U=\delta(t-t^*)$. Equivalently, finite values of $\P_i(t)$ indicate various degrees of spike repeatability. Of course, $\P_i(t)=0$ implies cell $i$ is not currently spiking on any trial.

\subsection{Spike events: repeatable temporal patterns}

Our next goal is to use $\P_i(t)$ to derive a metric of spike time reliability for a network. Intuitively, given a spike observed on one trial, we seek the expected probability that this spike would be present on any other trial. This amounts to asking to what extent the function $\P_i(t)$ is ``peaked"  on average.

To develop a practical assessment of this extent, we begin by approximating $\P_i(t)$ from a finite number of trajectories. To do so, we modify the definition of the flux from a continuous to a discrete time quantity. For practical reasons we say that $\P_i^{approx}(t)$ represents the fraction of a $\mu_\z^t$-ensemble of trajectories that crosses the $\o_i=1 \sim 0$ boundary within a small time interval $t+\D t$. As a discrete quantity, we now have $0 \leq \P_i^{approx}(t) \leq 1$. Borrowing a procedure from~\cite{Tiesinga:2008p12846}, we convolve this discretized flux  with a gaussian filter of standard deviation $\s$ to obtain a smooth waveform (see Fig~\ref{fig:event} (B)). We then define {\it spike events} as local maxima (peaks) of this waveform. A spike is assigned to an event if it falls within a tolerance window of the event time, defined by the width of the peak at half height. If the spikes contributing to an event are perfectly aligned, the tolerance is $\s$. However, if there is some variability in the spike times, the tolerance grows as the event's peak widens. This procedure ensures that spikes differing by negligible shifts are members of the same event. For our estimates, we used $\D t=0.005$ (time step of the numerical solver) and found that $\s=0.05$ was big enough to define reasonable event sizes and small enough to discriminate between most consecutive spikes from the same trial. However, we note that the following results are robust to moderate changes in $\s$.

Each spike event is then assigned a {\it participation fraction} $f$: the fraction of trials participating in the spike event.
Figure~\ref{fig:event} (B) shows the distribution of $f$'s for the events recorded from all cells of our chaotic network testbed, using 2500 time unit runs with 30 trials and discarding the initial 10\% to avoid transient effects. There is a significant fraction of events with $f=1$ and a monotonic decrease of occurrences with lesser participation fractions. The mean $\langle f \rangle$ of this distribution is the finite-sampling equivalent of the average height of  $\P_i$ peaks and therefore represents an estimate of the expected probability of an observed spike being repeated on other trials.

Finally, we compare $\langle f \rangle$ to the number of unstable directions of the chaotic attractor $\mu_\z^t$ for a range of input amplitude $\e$. Figure~\ref{fig:flux} (C) shows both $\e$-dependent curves $1-\langle f \rangle$ and $M_\l/N$ (previously shown in Fig~\ref{fig:color} (B)). For weak input amplitudes ($\e<0.2$), networks are asymptotically reliable and thus, $M_\l/N=0$ and every event has full participation fraction ($1-\langle f \rangle=0$). As $\e$ increases, the network undergoes a rapid transition from stable to chaotic dynamics. Most interestingly, both $1-\langle f \rangle$ and $M_\l/N$ follow the same trend, suggesting that the dimension of the underlying strange attractor plays an important role in the expected reliability of spikes. While this relationship is not perfect, it shows that the number of positive Lyapunov exponents serves as a better predictor of average spike reproducibility than the magnitude of $\l_1$ alone.

The shapes of $1-\langle f \rangle$ and $M_\l/N$ show an initial growth followed by a gradual decay, suggesting that following a transition from stable to chaotic dynamics, higher input fluctuations induce more reliable spiking. In the limit of high $\e$, this agrees with the intuition of an entraining effect by the input signal. 
This raises an important question about the observed dynamics: Is spike repeatability simply due to large deviations in the input? Or equivalently, is the role of chaotic network interactions comparable to ``noise" in the inputs to individual neurons?
That this may not be the case for moderate input amplitudes is suggested by the concentration of trajectories in the sample measures $\mu_\z^t$. We now seek to demonstrate the difference.

\section{Relevant local mechanisms}

\subsection{Network interactions vs. stimulus}

A natural question about the dynamical phenomena described above is: to what extent are they caused by network interactions, compared to direct effects of the stimulus?  In our system, each cell receives an external stimulus $\z_i(t)$  as well as a sum of inputs from other cells.  Because of network interactions, the latter inputs are highly structured even when $\l_1>0$, and can be correlated across multiple trials.  Indeed, all else being equal, the more singular and low-dimensional $\mu_\z^t$ is, the more cross-trial correlation there will be.  The question is whether we would still observe the same spiking behavior when inputs from the rest of the network are replaced by more random inputs.

To test this, we compare the response of cell $i$ in a network driven by the stimulus $\z(t)$ with that of a single ``test cell.''  The test cell receives:  (i) the $i$th component $\z_i(t)$ of the same stimulus, and (ii)  excitatory and inhibitory spike trains with statistics chosen to ``match" network activity in two different ways that we describe below.  For each, the number of such spike trains matches the mean in-degree $K$ of the network. That is, there are $K$ excitatory and $K$ inhibitory spike trains such balance is conserved.

In our first use of the test cell, we present poisson-distributed spike trains that are adjusted to the network firing rate (at each $\epsilon$).  Importantly, all trains are independent (both within and across trials). We denote the corresponding average spike event participation fractions by $\langle f^{poisson} \rangle$.

In our second use of the test cell, we present spike trains taken from $K$ excitatory and $K$ inhibitory cells, each chosen from a simulation with a different initial condition but with the {\it same} stimulus drive $\z(t)$.  This way, the stimulus modulation of the individual spike trains is preserved, but the global structure of the chaotic attractor is disrupted.  The corresponding average spike event participation fractions in this case are denoted by $\langle f^{shuffle} \rangle$.

Fig~\ref{fig:event} (C) shows $1-\langle f^{poisson} \rangle$ and $1-\langle f^{shuffle} \rangle$  alongside $1-\langle f \rangle$. For moderate values of $\e$, these three curves differ by a factor of 2 (poisson) and 1.5 (shuffle) and slowly converge as $\e$ increases. This confirms that two dynamical regimes are present: When the input strength is very high, inputs tend to entrain neurons into firing regardless of synaptic inputs, as was intuitively stated above. However, for moderate input amplitudes, network interactions play a central role in the repeatability of spike times. Importantly, we note that many repeatable spike events in chaotic networks are not present in the test cell driven with either surrogate poisson or trial-shuffled excitatory and inhibitory events, even though the same stimulus $\z_i(t)$ was given in each case.

A second, closely related question is whether the reliable spiking events we see are solely due to large fluctuations in the stimulus, or if network mechanisms play a significant role.  The above results, which show that structured network interactions can have a significant impact on single-cell reliability, suggest the answer is no.  Here we provide a second, more direct test of this question.  

To proceed, we first classify each spike fired in the network as either reliable or not by defining a quantity $\Rs$: the fraction of spikes belonging to an event with a participation fraction $f$ greater or equal to some threshold. $\Rs$ is the cumulative density of events with $f$ greater than the chosen threshold. Equivalently, we say a spike event is {\it reliable} if its $f$ is greater than that threshold and {\it unreliable} otherwise. Individual spikes inherit the reliability classification of the event of which they are a member.

For visual comparison with Fig~\ref{fig:flux} (C), Fig~\ref{fig:event} (D) shows $1-\Rs$ as a function of $\e$ for three threshold values (0.5, 0.75 and 1). These curves show the fraction of unreliable spikes, out of all spikes fired, for a given threshold. The error bars show the standard deviation of the value across all cells in the network. As expected for small $\e$, $1-\Rs=0$ since $\l_1<0$. Notice that as in the case of $1-\langle f \rangle$, the distinct choices of threshold do not affect overall trends, but they greatly impact the fraction of spikes labeled reliable (or unreliable).
%
For what follows, we adopt a strict definition of spike time reliability by fixing the $\Rs$ threshold at 1 (i.e. a spike is reliable if it is present in all trials). However, the subsequent results are fairly robust to the choice of this threshold. 

We can now address the question raised above via {\it spike-triggered averaging} (STA).  As the name describes, this procedure takes quantities related to a given cell's dynamics (i.e. stimulus, synaptic inputs, etc.) in the moments leading to a spike, and averages them across an ensemble of spike times. In other words, it is a conditional expectation of the stimulus in the moments leading up to a spike; it can also be interpreted as the leading term of a Wiener-Volterra expansion of the neural response \cite{spikesbook}.   In what follows, we will distinguish between reliable and unreliable spikes while taking these averages in an effort to isolate dynamical differences between the two.

\begin{figure}
\begin{center}
\includegraphics{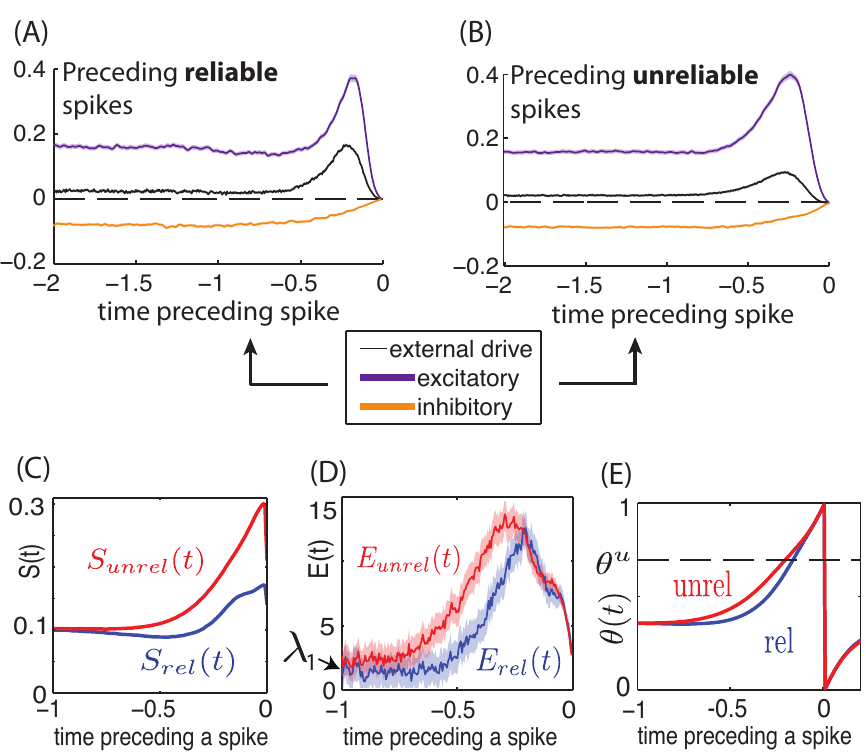}
\caption {(Color online) For (A) through (E), $t=0$ marks the spike time and rel/unrel indicates the identity of the spike used in the average. (A) and (B), Spike triggered averaged external signal $\e Z(\o_i)\z_i(t)$ (black), excitatory (purple) and inhibitory (orange) network inputs $Z(\o_i)\sum_j a_{ij}g(\o_j)$. (A) Triggered on reliable spikes. (B) Triggered on unreliable spikes.  (C) Spike triggered support score $S$. (D) Spike triggered local expansion measure $E$. (E) Spike triggered average phase $\o_i$. For all panels: $\n=-0.5$, $\e=0.5$ with $\l_1 \approx 2.5$. Shaded areas surrounding the computed averages show two standard errors of the mean.\footnote{Computed standard deviations where verified by spot checks using the method of batched means with about 100 batches of size 1000.}  No shade indicates that the error is too small to visualize.}
\label{fig:expansion1}
\end{center}
\end{figure}%

For illustration, we turn to our chaotic network testbed.
Figure~\ref{fig:expansion1} (A) and (B) shows the STA of both excitatory and inhibitory network interactions as well as the external input leading to reliable and unreliable spikes. More precisely, say we consider spike times $\{ t_i^1, t_i^2, ... \}$ from cell $i$. Then the network interactions used in the STA is the ensemble of time traces $\{Z(\o_i(t))\sum_j a_{ij}g(\o_j(t)) | t_i^*-2 \leq t \leq t_i^*\}$ where $Z(\o)$ and $g(\o)$ are as in~(\ref{stoch_tht}) where we differentiate between excitatory and inhibitory inputs according to the sign of $a_{ij}$. Similarly, the external input are taken from $\{ \e Z(\o_i(t))\z_i(t) | t_i^*-2 \leq t \leq t_i^*\}$.

There are two main points to take from these STAs.  First, note that the {\it average} levels of spike-triggered recurrent excitation and inhibition very roughly balance one another in time periods well before spike times.  However, right before spikes this balance is broken, leading to an excess of recurrent excitation which is stronger than the spike-triggered stimulus.  This gives further evidence that recurrent interactions shape the dynamics with which the spikes themselves are elicited --- rather than spikes being primarily driven by the external stimuli alone.  Second, note that these STAs are qualitatively similar for both reliable and unreliable spikes. Even though the peak of the summed external input in Fig.~\ref{fig:expansion1}(A) is higher than that in Fig.~\ref{fig:expansion1}(B), it is not clear that this difference is sufficient, by itself, to explain the increase in reliability (as the magnitude of the mean external input is relatively small).  This suggests that we look
other dynamical factors that might contribute to reliable spike events, a task to which we now turn.

Recall that the support score $s_i(t)$ measures the contribution of a single cell's subspace to tangent unstable directions of a trajectory. Consider the corresponding STA $S(t)$, i.e., the expected values of $s_i(t)$ in a short time interval preceding each spike in the network. Fig.~\ref{fig:expansion1} (C) shows the resulting averages for both reliable and unreliable spikes. Moments before a cell fires an unreliable spike, $S(t)$  is considerably larger than in the reliable spike case, thus indicating that global expansion is further aligned with a spiking cell's direction in unreliable spike events. We now investigate properties of the flow leading to this phenomenon.

\subsection{Source of local expansion}

To better capture space expansion in a given neural direction, consider $v(t)$, the solution of the variational equation 
\beq
\dot{v}=J(t)v
\label{var_max}
\eeq
where $J(t) = D\Psi_{0,t;\z}$ is the Jacobian of the flow evaluated along a trajectory $\o(t)$. If we set $v(0)$ to be randomly chosen but with unit length, then $v(t)$ quickly aligns to the directions of maximum expansion in the tangent space of the flow about $\o(t)$; moreover, because of ergodicity 
$
\l_1=\lim_{t \to \infty} \frac{1}{t}\log(\|v(t)\|).
$
We can equivalently write a discretized version of this expression for small $\D t$: $\l_1=\lim_{T \to \infty} \langle e(t) \rangle_T$ where $\langle \cdot \rangle_T$ denotes the time average up to time $T$ and
$
e(t)=\frac{1}{\D t}\log\left(\frac{\|v(t+\D t)\|}{\|v(t)\|}\right)
$
is analogous to a finite time Lyapunov exponent.  For our network, $e(t)$ fluctuates rapidly and depends on many factors such as number of spikes fired, the pattern of the inputs, and the phase coordinate of each cell over the time $\D t$. Its coefficient of variation is typically $\O(10)$ for $\D t =0.005$ which is consistent with the fact that stability is very heterogeneous in time. 
To better understand the behavior of the flow along single neural directions, we define the {\it local expansion coefficient}
\beq
\begin{split}
e_i(t)&=\frac{1}{\D t}\log \left( \frac{|v_i(t+\D t)|}{|v_i(t)|} \right).
\label{metrics}
\end{split}
\eeq%
Note that $e_i(t)$ is a local equivalent of $e(t)$ and directly measures the maximum expansion along a neural direction.
 
Define $E(t)$ as the STA corresponding to $e_i(t)$, shown in Fig.~\ref{fig:expansion1}  (D). Notice that at its peak, $E_{\text{unrel}}(t)$ is much broader than $E_{\text{rel}}(t)$, with 
$
\int_{-2}^{0} E_{\text{unrel}}(t)-E_{\text{rel}}(t)dt \simeq 2.5
$
which indicates that prior to an unreliable spike, trajectories are subject to an accumulated infinitesimal expansion rate higher than in the reliable spike case. 

In contrast to $s_i(t)$, $e_i(t)$ is directly computable in terms of contributions from different terms in the flow. We refer the reader to the Appendix for a detailed treatment of input conditions leading to reliable or unreliable spikes. Importantly, the source of ``local" expansion $e_i(t)$ is dominated by the effect of a single cell's vector field $F(\o_i)$ (from Eqn.~(\ref{stoch_tht})) which directly depends the phase trajectory $\o_i(t)$ prior to a spike.

 If $\o_i(t)<\frac{1}{2}$, $F'(\o_i(t))$ is negative, and becomes positive for $\o_i(t)>\frac{1}{2}$ --- in absence of fluctuating inputs from network or external source). When an uncoupled cell is driven by $\z_i$, we know that on average, it spends more time in its contractive region ($\o_i<\frac{1}{2}$) and is reliable as a result~\cite{LinSY07a, Ritt:2003p145}. While inputs may directly contribute to $J(t)$, their effect is generally so brief that their chief contribution to $e_i(t)$ is to steer $\o_i(t)$ toward expanding regions of its own subspace (see Appendix). Fig~\ref{fig:expansion1} (C) confirms that the average phase of a cell preceding an unreliable spike spends more time in its expanding region. Such a phenomenon has previously been reported in the form of a threshold crossing velocity argument~\cite{Banerjee:2006p16587}.

The key feature of this driven system, likely due to sparse and rapid coupling, is a sustained balance between inputs leading to contraction/expansion in local neural subspaces. A bias toward more occurrences of ``expansive inputs" yields positive Lyapunov exponents ($M_\l>0$) and implies on average, more growth than decay. What is perhaps surprising is that this state space expansion remains confined to subspaces supported by only a few neural directions, which creates this coexistence of chaos and highly reliable spiking throughout the network.

\section{Discussion} 
In this article, we explored the reliability of fluctuation-driven networks in the excitable regime --- where model single cell dynamics contain stable fixed points. We showed that these networks can operate in stable or chaotic regimes
and demonstrated that spike trains of single neurons from chaotic networks can retain a great deal of temporal structure across trials. 
We have found that an attribute of random attractors that directly impacts the reliability of single cells is the orientation of expanding subspaces, and that the evolving shape of the random attractor is reflected in the intermittent reliability of single neurons. 
We have also performed a detailed numerical study to analyze the local (i.e., cell-to-cell) interactions responsible for reliable spike events.

This said, a mechanistic understanding of the origins of chaotic, structured spiking remains to be fully developed.  Specifically, we still need to work out the role of larger-scale network structures, and how unreliable spike events propagate through the network in a self-sustaining fashion in networks with $\l_1>0~.$  This is a target of our future work.

Throughout this work, we have found the qualitative theory of random dynamical systems to be a useful conceptual framework for studying reliability. Though the theory is predicated on a number of idealizations, we expect most of them (e.g., the assumption that the stimuli are white noise rather than some other type of stochastic process) can be relaxed.

Finally, we note that the phenomena observed here may have consequences for neural information coding and processing.  In particular, unreliable spikes are a hallmark of sensitivity to initial conditions and may therefore carry information about previous states of the system (or, equivalently, previous inputs).  In contrast, reliable spikes carry repeatable information and computations  about the external stimulus $\z(t)$ (either via directly evoked spikes or propagated by repeatable network interactions). 
We showed that both unreliable and reliable spike events coexist in chaotic regimes of the system explored. Preliminary results indicate that correlation across external drives greatly enhances a network's spike time reliability and will be the object of an upcoming publication.
 The resulting implications for the neural encoding of signals are an intriguing avenue for further investigation.

 \section{Acknowledgments}
The authors thank Lai-Sang Young for helpful insights. This work was supported in part by an NSERC graduate scholarship, an NIH Training Grant, the Burroughs Wellcome Fund Scientific Interfaces, and the NSF under grant DMS-0907927. Numerical simulations performed on NSF's XSEDE supercomputing platform.

\appendix
\section{APPENDIX}

\section{Model and coordinate transformations from QIF}
Our networks are composed of {\it $\o$-neurons}, which are equivalent to the quadratic-integrate-and-fire (QIF) model~\cite{LathamRNN00,Ermentrout:1996p10447}. The latter is formulated in terms of membrane potentials, and thus has a direct physical interpretation. However, it is a hybrid dynamical system, i.e., its solutions are instantaneously reset to a base value after a spike is emitted.  For our purposes, such discontinuities are rather inconvenient. Fortunately, there exists a smooth change of coordinates mapping the QIF (hybrid) dynamics to the $\o$-space, where a cell's membrane potential is represented by a phase variable on the unit circle $S^1$. This representation has the advantage of being one of the simplest to capture the nonlinear spike generating mechanisms of Type I neurons with solutions that remain smooth and live on a compact domain, a mathematical feature central to this study. We now review this change of coordinates and the equivalence of the two models.

The variable $v$ represents the membrane potential of a single neuron and its dynamics are described by the following equation:
\beq
\tau \frac{dv}{dt}=\frac{(v-v_R)(v-v_T)}{\D v} +I_a+I_d(t)
\label{basic_qif}
\eeq
where $\tau$ is the cell membrane time constant, $v_R$ and $v_T$ are rest and threshold voltages respectively and $\D v=v_T-v_R$. $I_a$ is an applied constant current, and $I_d(t)$ is a time varying input drive. If $v(t)$ crosses the threshold $v_T$, its trajectory quickly blows up to infinity where it is said to fire a spike. Once a spike is fired, $v(t)$ is reset to $-\infty$ and the trajectory will converge toward $v_R$. To implement this in simulations, a ceiling value is set such that when reached, it represents the apex of a spike and the voltage $v(t)$ is ``manually" reset to a value below threshold. As we will see, the $\o$-model circumvents the need for this procedure.

In absence of other inputs ($I_d=0$), the baseline current $$I_a=I^*=\frac{(v_T-v_R)^2}{4\D v}=\frac{\D v}{4}$$ places the system at a saddle node bifurcation, responsible for the onset of tonic (periodic) firing. Therefore, if $I_a<I^*$, the neuron is said to be in {\it excitable regime} whereas if $I_a>I^*$, it is in {\it oscillatory regime}.

Let us suppose that the input term $I_d(t)$ is a realization of a white noise process scaled by a constant $\rho$. We can rewrite~(\ref{basic_qif}) as a stochastic differential equation (SDE)
\beq 
\tau d v= \left ( \frac{(v-v_R)(v-v_T)}{\D v} +I_a \right ) dt+\rho dW_t
\label{stoch_qif}
\eeq
where $W_t$ is a standard Wiener process. We treat~(\ref{stoch_qif}) as an SDE of the It\^{o} type~\cite{Lindner:2003p16585} as it is more convenient for numerical simulations and carry out the change of variables accordingly.

Let us introduce a new variable $\o$ defined by
\beq
v(\o)=\frac{v_T+v_R}{2}+\frac{\D v}{2}\tan((2\pi\o-\pi)/2)
\label{v_tht}
\eeq
along with a rescaling of time 
\beq
t \mapsto \frac{t}{4\pi\tau}.
\label{time}
\eeq
Equation~(\ref{stoch_qif}) now reads
\beq
\begin{split}
 d\o&=\left[ F(\o)+\n Z(\o)+ \frac{\e^2}{2} Z(\o)Z'(\o)\right] dt+\e Z(\o)dW_{t}
\end{split}
\label{stoch_tht2}
\eeq
 where $F(\o)=1+\cos(2\pi \o)$, $Z(\o)=1-\cos(2 \pi \o)$ and
$$\n=\frac{4}{\D v}I_a-1$$
$$\e=\frac{2\rho}{\D v\sqrt{\tau\pi}}$$ which is the $\o$-model on $[0,1]$ we want. In absence of stochastic drive and for $\n<0$, the two fixed points are given by
$$\o^{s,u}=\frac{1}{2\pi}\arccos(\frac{\n+1}{\n-1})$$
which for $\n=-1$ yields $\o^s=1/4$ and $\o^u=3/4$. For $\n>0$, the neuron fires periodically at a frequency of $\sqrt{\n}/2$.

Typical parameter choices for the QIF model are
\beq
\begin{split}
\tau&=10ms\\
v_R&=-65mV\\
v_T&=-50mV
\end{split}
\label{qif_params}
\eeq
with time in units of milliseconds. Expression~(\ref{time}) implies that one time unit in $\o$-coordinates corresponds to about $125$ milliseconds. In the absence of applied current $I_a$, we get $\eta=-1$.

\subsection{Network architecture and synaptic coupling}

In the main manuscript, we explore the dynamics of  Erd\"{o}s-Renyi type random networks of $N=1000$ cells of which 80\% are excitatory and 20\% inhibitory. Each cell receives on average $K=20$ synaptic connections from each excitatory and inhibitory subpopulation. We implement a classical balanced state architecture and scale synaptic weights of these connections by $1/\sqrt{K}$ which ensures that fluctuations from network interactions remain independent of $K$ in the large $N$ limit~\cite{Vreeswijk:1998p14451} (as long as $K<<N$ and cells fire close to independently). Although we do not systematically explore the scaling effects of $N$ and $K$, preliminary results for combinations of $K=$ 50, 100, 200 and $N=$2500, 5000 indicate that our findings are qualitatively robust to system size. 

As mentioned earlier, one of the advantages of the $\o$-neuron model is the continuity of dynamics in phase space. We can therefore easily implement synaptic interaction between two neurons with differentiable and bounded terms. 
Synaptic interactions between $\o$-neurons are modeled using a smooth function 
 \beq
g(\o)=\left\{\begin{array}{cl}d \left( b^2 - \left[\left(\o+\frac{1}{2}\right)\text{mod}\,\, 1-\frac{1}{2}\right]^2 \right)^3 & \text{; }\o \in [-b,b] \\
0 & \text{; else}
\end{array}\right.
\eeq
where $b = \frac{1}{20}$ and $d=\frac{35}{32}$.  

For example, for two cells coupled as $2 \rightarrow 1$, we have 
\beq
\dot{\o}_1=F(\o_1)+Z(\o_1)\left( \n+a_{12}g(\o_2)\right)
\label{tht_net}
\eeq
in which $a_{12}$ is the synaptic strength from neuron $2$ to neuron $1$.
Neuron 2 only affects $\o_1$ when  $\o_2\in[-b,b]$, mimicking a rapid rise and fall of a synaptic variable in response to presynaptic potential fluctuation during spike generation.
We follow the approach of Latham et al.~\cite{LathamRNN00}  to assess the effective coupling strength from neuron $\o_2$ to neuron $\o_1$ in the form of evoked post synaptic potentials (PSP).  Specifically, we first derive a relationship between the value of $a_{12}$ and the evoked PSP following a presynaptic spike from $\o_2$ in the $\theta$ coordinates.    We then translate this to the voltage coordinates. 

We assume that $\n=-1$ (equiv. to $I_a=0$), and that $\o_1$ sits at rest $\o_R=\frac{1}{4}$, and compute the value $\o_S=\o_R+\o_{PSP}$.
As the support of $g$ is quite small, let us linearize~(\ref{tht_net}) for $\o_2$ when it crosses $0\sim1$. We obtain neuron 2's phase velocity at spike time, $\dot{\o}_2=2$, and hold this velocity constant in the calculation that follows. Suppose that at $t=0$, $\o_2$ is at the left end of $g$'s support, then
\beqn
\o_2(t)=2t-b
\eeqn
which gives us the non-autonomous equation for $\o_1$
\beq
\dot{\o}_1=F(\o_1)+Z(\o_1)[-1 +a_{12}g(2t-b)] \quad , \quad \o_1(0)=\o_R=1/4.
\label{psp}
\eeq
We  make a final assumption for small PSPs and assume that the behavior of~\eqref{psp} is linear about the resting phase ($\o_1=\o_R$). This yields
$
\dot{\o}_1=a_{12}g(2t-\b)
$
which in turn gives us
\beqn
\int_{\o_R}^{\o_S}d\o_1=a_{12}\int_0^{t=\b}g(2t-\b)dt
\eeqn
Notice that $\int_{-b}^{b}g(\o)d\o=1$, which gives the relationship
\beqn
a_{12}=2(\o_S-\o_R).
\eeqn

Although we have made fairly strong assumptions about the $\o$-dynamics in deriving this expression, we tested it numerically and found that predictions of post synaptic $\o$ variations were accurate up to the third significant digit, for the range of PSPs of interest. Using~(\ref{v_tht}), we get the equivalent expression:
\beqn
v_{PSP}=\frac{v_T+v_R}{2}+\frac{\D v}{2}\tan((\pi a_{12}-\pi)/2).
\eeqn
For $K=20$ and $a_{ij}=1/\sqrt{K}$, we get the following approximations for excitatory and inhibitory PSPs: $v_{EPSP} \simeq 4.0mV$ and $v_{IPSP} \simeq -8.4 mV$.

Finally, we note that all synaptic couplings, when present between two cells, are of the same strength throughout the network (only the sign changes to distinguish between excitatory and inhibitory connections). Additionally, while  both $\n$ and $\e$ are network-wide constants, we introduce $\O(10^{-2})$ perturbations randomly chosen for each cell in order to avoid symmetries in the system.

\section{Lyapunov spectrum approximation}

In the main manuscript, we present approximations of the Lyapunov spectrum  $\l_1\geq\l_2\geq ... \geq \l_N$ and related quantities for the network described by~(\ref{stoch_tht}).
Under very general conditions, the $\l_i$ are well defined for system (\ref{stoch_tht}) and that they do not depend on the choice of IC or $\z(t)$. 
However, the Lyapunov exponents generally cannot be computed analytically and we therefore use Monte-Carlo simulations to approximate them. We numerically simulate system (\ref{stoch_tht}) using a Euler-Maruyama scheme with time steps of 0.005. At each point in time, we simultaneously solve the corresponding variational equation 
\beq
\dot{S}=J(t)S
\label{var_comp}
\eeq
where $J(t)$ is the Jacobian of the flow evaluated along the simulated trajectory and $S(0)$ is the $N \times N$ identity matrix. The solution matrix $S(t)$ is then orthogonalized at each time step in order to extract the exponential growth rates associated with each Lyapunov subspace. See~\cite{Geist:1990p12843} for details of this standard algorithm.

All reported values of $\l_i$ have a standard error less than 0.002, estimated by the method of batched means~\cite{Asmussen2007} (batch size = 500 time units) and cross-checked using several realizations of white noise processes and random connectivity matrices. We have also verified, by spot checks, that varying the batch window size does not affect the error estimate significantly. 

Numerical simulations were implemented in Python and Cython programming languages and carried out on NSF's XSEDE supercomputing platform. 

\section{Spike triggered flow decomposition}
\begin{figure}
\begin{center}
\includegraphics{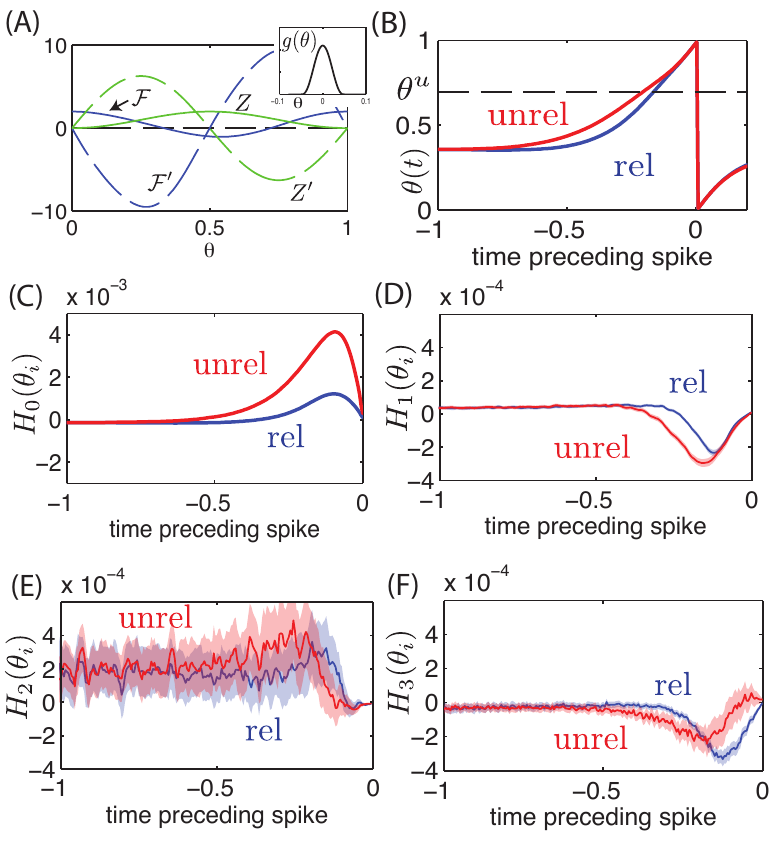}
\caption{(Colors online) (A) Distinct terms of the single cell flow and Jacobian. Inset :  synaptic coupling function $g(o)$. For panels (B-F), $t=0$ marks the spike time and rel/unrel indicate the identity of the spike used in the average. (B) Spike triggered average phase $\o_i$ (same as in Fig~\ref{fig:expansion1} E). (C-F) Spike triggered average terms $H_0(\o_i)$, $H_1(\o_i)$, $H_2(\o_i)$ and $H_3(\o_i)$. Network parameters: $\n=-0.5$, $\e=0.5$, yielding $\l_1 \simeq 2.5$. For all panels except (A): shaded areas surrounding the computed averages show two standard errors of the mean.\footnote{ Standard errors of the mean were verified via spot checks using the method of batched means,  with about 100 batches of size 1000.} No shade indicates that the error is too small to visualize.} 
\label{fig:expansion2}
\end{center}
\end{figure}
We take a closer look at the single-cell flow in an attempt to better understand the origin of reliable and unreliable spikes. We concentrate on the effect of inputs on the local expansion coefficient $e_i(t)$ (see Eqn.~(\ref{metrics})). Consider the time evolution of $v_i(t)$ by unpacking the $i^{\text{th}}$ component of the discretized version of~(\ref{var_max}):
\beq
\begin{split}
&v_i(t+\D t)=v_i(t)+ v_i(t)[\D t\F(\o_i)+\D t Z'(\o_i)\sum_j a_{ij}g(\o_j)\\
&+\sqrt{\D t} \xi_t \e Z'(\o_i)] +\D t Z(\o_i)\sum_j a_{ij}g'(\o_j)v_j(t)
\label{unpack}
\end{split}
\eeq
where $\F(\o_i)=F(\o_i)+\n Z(\o_i)+\frac{\e^2}{2}Z(\o_i)Z'(\o_i)$ and $\xi_t\sim N(0,1)$; $\D t$ is the time  increment. 

We substitute expression~(\ref{unpack}) as the numerator in the definition of $e_i(t)$ (Eqn.~(\ref{metrics})), in order to discern the contribution of different terms in the network dynamics to state space expansion. 
Let us define the following terms 
\beq
\begin{split}
H_0(t)=& \D t |v_i(t)|\F'(\o_i)\\
H_1(t)=& \D t |v_i(t)|Z'(\o_i(t))\sum_j a_{ij}g(\o_j(t))\\
H_2(t)=& \D t Z(\o_i(t))\sum_j a_{ij}g'(\o_j(t))|v_j(t)|sgn[v_i(t)v_j(t)]\\
H_3(t)=& \sqrt{\D t} |v_i(t)| \e Z'(\o_i(t)) \xi_t.
\end{split}
\label{terms}
\eeq
Notice the use of the absolute value for $v(t)$ components which ensures that $H_k(t)>0$ implies expansion (or, if $H_k(t)<0$, contraction) in whichever of the positive and negative directions $v_i(t)$ is pointing. 

Here, $H_0$ captures the contribution of the single-cell vector field to the proportional growth (or decay) of $v_i(t)$.  Note from Fig~\ref{fig:expansion2} (A) that $\F'(\o_i)$, the main contributing part of $H_0$, is negative for $\o_i \in (0,1/2)$ and positive for $\o_i \in (1/2,1)$. Meanwhile, $H_1$ measures the contribution of synaptic inputs and $H_2(t)$ the relative contribution of presynaptic neurons' coordinates. The latter varies quite rapidly, because the derivative of the coupling function $g(\o_j)$ (shown in Fig~\ref{fig:expansion2} (A)) takes large positive and negative values. In essence, it quantifies the transfer of expansion from one cell to the next: if $|v_j(t)|$ is large, and $sgn[v_i(t)v_j(t)]=1$, then $H_2$ causes expansion in the $v_i(t)$. Finally, $H_3$ captures the contribution of the external drive. 

We now assess the relative importance of all of these dynamical effects to spike time reliability. We do this by comparing the magnitude and sign of the $H$ terms.  Specifically, we compute spike-triggered averages of these terms in periods before reliable and unreliable spike events.  We continue to use the criterion from the main text that a spike is considered reliable if it occurs on each of the simulated trials.

Notice first that inputs -- synaptic or external -- enter multiplicatively with $Z'(\o_i)$, which is negative in $\o \in (\frac{1}{2},1)$ (see Fig~\ref{fig:expansion2}). This implies that more-excitatory synaptic inputs -- or more-positive external inputs -- arriving shortly before spikes promote contraction for $H_1$ and $H_3$.  We see that both of these terms are primarily negative in the time periods before spikes (Panels (D),(F)), as positive inputs push cells across the spiking threshold.  

Expansion -- especially for unreliable events -- arises from the coupling term $H_2$ and from the term $H_0$ representing internal dynamics.  Note in particular that this latter term is an order of magnitude higher than the others; thus, we focus our attention on this  next.  Panel (B) of Fig~\ref{fig:expansion2} shows that the speed at which phases cross the threshold is lower for unreliable spikes than for reliable ones.  This further explains why the $H_0$ averages --mainly depending on $\F'$-- are larger for the unreliable spikes.

Thus, we conclude -- as noted in the main text -- that the primary dynamical mechanism behind the unstable dynamics is that inputs steer $\o_i(t)$ in expansive regions of its own subspace (see Fig~\ref{fig:expansion2} (C) or Fig~\ref{fig:expansion1} (E)). 
This conclusion that instabilities in the flow are mainly generated by intrinsic dynamics is interesting, as it suggests that network stability could vary in rich ways depending on cell type and spike generation mechanisms.

 \begin{small}
 \bibliographystyle{plain}
%

 \end{small}

\end{document}